\documentclass[Physsubmission, Phys]{SciPost}

\hypersetup{
    colorlinks,
    linkcolor={red!50!black},
    citecolor={blue!50!black},
    urlcolor={blue!80!black}
}

\usepackage{booktabs} 
\usepackage[bitstream-charter]{mathdesign}
\DeclareSymbolFont{usualmathcal}{OMS}{cmsy}{m}{n}
\DeclareSymbolFontAlphabet{\mathcal}{usualmathcal}
\newcommand{\spazio}{~~~~~~~~~~}
\newcommand{\hspazio}{~~~~~}

\begin{document}

\begin{center}{\Large \textbf{
NLO QCD corrections for off-shell $t\bar{t}b\bar{b}$\\
}}\end{center}

\begin{center}
Michele Lupattelli\textsuperscript{1$\star$}
\end{center}

\begin{center}
{\bf 1} Institute for Theoretical Particle Physics and Cosmology, RWTH Aachen University,
D-52056 Aachen, Germany
\\
* lupattelli@physik.rwth-aachen.de
\end{center}

\begin{center}
\today
\end{center}


\definecolor{palegray}{gray}{0.95}
\begin{center}
\colorbox{palegray}{
  \begin{tabular}{rr}
  \begin{minipage}{0.1\textwidth}
    \includegraphics[width=35mm]{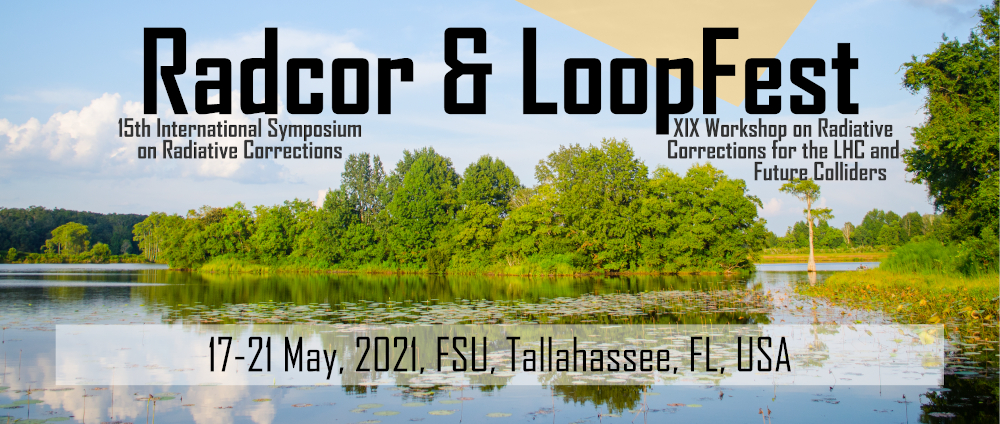}
  \end{minipage}
  &
  \begin{minipage}{0.85\textwidth}
    \begin{center}
    {\it 15th International Symposium on Radiative Corrections: \\Applications of Quantum Field Theory to Phenomenology,}\\
    {\it FSU, Tallahasse, FL, USA, 17-21 May 2021} \\
    \doi{10.21468/SciPostPhysProc.?}\\
    \end{center}
  \end{minipage}
\end{tabular}
}
\end{center}

\section*{Abstract}
{\bf
We present a full off-shell NLO QCD calculation of $pp \rightarrow e^+ \nu_e b \mu^- \bar{\nu}_\mu \bar{b} b \bar{b} + X$. Results are obtained using the \textsc{Helac-Nlo} Monte Carlo framework. We discuss a comparison to previous results and to experimental measurements. We also investigate the contribution of the initial states involving $b$-quarks and we introduce two $b$-jet tagging schemes.
}

\vspace{10pt}
\noindent\rule{\textwidth}{1pt}
\tableofcontents\thispagestyle{fancy}
\noindent\rule{\textwidth}{1pt}
\vspace{10pt}

  \section{Introduction}
  In 2012 the Higgs boson has been first observed at the Large Hadron Collider (LHC). Since then, tests are ongoing on the Higgs sector, the aim of which is to fully understand its properties. The Higgs boson couples to the fundamental fermions via the Yukawa interaction. Among them, the heaviest one is the top-quark. A direct probing of the coupling of the Higgs boson to the top-quark, the top-Yukawa coupling $Y_t$, is provided by the process $pp\rightarrow t\bar{t}H$, which represents $1\%$ of the total Higgs boson cross section. Because of its mass of about $m_H = 125$ GeV, the Higgs boson mainly decays into a bottom-quark pair, with a branching ratio of $58\%$ \cite{LHCHiggsCrossSectionWorkingGroup:2016ypw}. Therefore $pp\rightarrow t\bar{t}H \rightarrow t\bar{t}b\bar{b}$ is a prime ingredient to extract information on $Y_t$. But this channel is very challenging because of the so-called \textit{combinatorial background}. Indeed, because of the presence of a top-quark pair, after the decays the final state presents $4$ $b$-jets. The same final state can be achieved by the direct $t\bar{t}b\bar{b}$ production. From an experimental point of view, light jets can be misidentified as $b$-jets. Hence, also $t\bar{t}jj$ production needs to be taken into account. To sum up, for a complete description of $t\bar{t}H$ we need to study accurately the following processes:
\begin{align*}
  \text{Actual signal: } pp & \rightarrow t\bar{t}H \rightarrow t\bar{t}b\bar{b} \rightarrow W^+ W^- b\bar{b}b\bar{b} \\
  \text{Irreducible background: } pp & \rightarrow t\bar{t}b\bar{b} \rightarrow W^+ W^- b\bar{b}b\bar{b} \\
  \text{Reducible background: } pp & \rightarrow t\bar{t}jj \rightarrow W^+ W^- b\bar{b}jj
\end{align*}
In this proceeding, we are going to focus on the irreducible background. In particular, we provide full off-shell NLO QCD predictions at the integrated and differential level in the dileptonic final state $pp \rightarrow e^+ \nu_e b \mu^- \bar{\nu}_\mu \bar{b} b \bar{b} + X$. More details on this calculation can be found in \cite{Bevilacqua:2021cit}. We first discuss the setup of the calculation (sec.~\ref{sec:setup}). Then, we present results at the integrated (sec.~\ref{sec:integr}) and differential level (sec.~\ref{sec:diff}). Moreover, we present a comparison of our results to the ones already present in the literature \cite{Denner:2020orv} and to the experimental measurements \cite{ATLAS:2018fwl}. Then, we show that the contributions involving $b$-quarks in the initial state are negligible (sec.~\ref{sec:inib}). There, we introduce two different $b$-jet tagging schemes: one that takes into account the electric charge of the $b$-jet (\textit{charge aware}), and one that does not (\textit{charge blind}). Finally, we summarise our results and give an outlook on the future works (sec.~\ref{sec:summ}).
  
  \section{Setup of the calculation}
  \label{sec:setup}
\begin{figure}
\centering
{\includegraphics[width=0.9\textwidth]{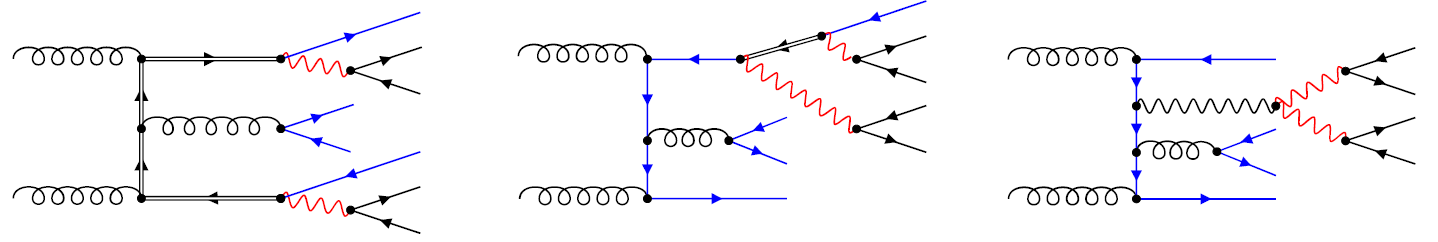}}
\caption{Representative Feynman diagrams for double- single- and non-resonant contributions to $pp \rightarrow e^+ \nu_e b \mu^- \bar{\nu}_\mu \bar{b} b \bar{b} +X$.}
\label{fig:res}
\end{figure}

  NLO QCD calculations of $t\bar{t}b\bar{b}$ with on-shell top-quarks are available by some time \cite{Bredenstein:2008zb, Bredenstein:2009aj, Bevilacqua:2009zn, Bredenstein:2010rs, Worek:2011rd, Bevilacqua:2014qfa}. These predictions can give a general idea about the size of the NLO QCD corrections, but cannot provide a reliable description of the top-quark decay products and the radiation pattern. The first step in this direction was done by matching these kind of predictions to parton shower algorithms \cite{Kardos:2013vxa, Cascioli:2013era, Garzelli:2014aba, Bevilacqua:2017cru}. These calculations provide information on the radiation pattern, but either the top-quark decays are omitted or performed in the parton shower. More recently first results with LO spin correlations in top-quark decays came out \cite{Jezo:2018yaf}. The first full off-shell calculation dates back only to last year and it was performed in the dileptonic decay channel \cite{Denner:2020orv}. In this proceeding we present the calculation for the same process performed using the \textsc{Helac-Nlo} Monte Carlo framework \cite{Bevilacqua:2011xh}. Namely, we computed NLO QCD corrections to $pp \rightarrow e^+ \nu_e b \mu^- \bar{\nu}_\mu \bar{b} b \bar{b} +X$ for LHC center of mass energy of $\sqrt{s}=13$ TeV. The $5$ flavour scheme is employed. In this calculation we take into account all the full off-shell effects, which means that the off-shell top-quarks are described by Breit-Wigner propagators, that we include all the double-, single- and non-resonant contribution (Fig.~\ref{fig:res}) and all the interference effect are consistently incorporated at the matrix element level.
Therefore, this is a complex calculation. We used \textsc{Helac-1Loop} \cite{vanHameren:2009dr}, \textsc{CutTools} \cite{Ossola:2006us, Ossola:2007ax} and \textsc{OneLOop} \cite{vanHameren:2010cp} to compute the virtual contribution. The most complicated one-loop diagrams in our calculation are octagon-type. The real corrections are computed with \textsc{Helac-Dipoles} \cite{Czakon:2009ss} using two different subtraction schemes: the Catani-Seymour subtraction scheme \cite{Catani:1996vz, Catani:2002hc} and the Nagy-Soper \cite{Bevilacqua:2013iha} subtraction scheme. This allowed us to check the correctness of the real contribution in an even more robust way than just relying on the internal consistency checks of each subtraction scheme \cite{Bevilacqua:2009zn, Nagy:1998bb, Nagy:2003tz, Czakon:2015cla}. Because of the complex nature of this calculation, we stored our theoretical predictions in the form of modified Les Houches Files \cite{Alwall:2006yp} and ROOT Ntuples \cite{Antcheva:2009zz, Bern:2013zja}. Thanks to this, we can change kinematical cuts, renormalization and factorization scales, PDFs and define new observables without the need of rerunning the whole calculation. Moreover, to save memory usage, we generated unweighted events.

In the following we describe our default setup. The default cuts are:
\begin{equation}
p_T(\ell) > 20~\text{GeV}, \hspazio |y(\ell)| < 2.5, \hspazio p_T(b) > 25~\text{GeV}, \hspazio |y(b)| < 2.5, \hspazio \Delta R(bb) > 0.4.
\end{equation}
We set the factorization scale equal to the renormalization scale $\mu_R = \mu_F$. We use two default scales: a fixed one $\mu_0 = m_t$ and a dynamical one $\mu_0 = H_T/3$, where $H_T$ is defined as follows
\begin{equation}
H_T = p_T(b_1) + p_T(b_2) + p_T(b_3) + p_T(b_4) + p_T(e^+) + p_T(\mu^-) + p_T^{miss}. 
\end{equation}
Our default PDF set is NNPDF3.1 \cite{NNPDF:2017mvq}. We use the LO set for the LO calculation and the NLO set for the NLO calculation. The results presented in section~\ref{sec:integr} and \ref{sec:diff} do not contain contributions from initial states involving $b$-quark. We assumed them to be negligible. A study on these initial state contributions is presented in section~\ref{sec:inib}.
  
  \section{Integrated fiducial cross sections}
  \label{sec:integr}
  We now present the results of our calculation. Using our default settings and the fixed scale, the integrated fiducial cross section of $pp \rightarrow e^+ \nu_e b \mu^- \bar{\nu}_\mu \bar{b} b \bar{b} + X$ is
\begin{align}
\sigma^{\text{LO}} & = 6.998^{+4.525(65\%)}_{-2.569(37\%)}\text{[scales] fb} \\
\sigma^{\text{NLO}} & = 13.24^{+2.33(18\%)}_{-2.89(22\%)}\text{[scales]}^{+0.19(1\%)}_{-0.19(1\%)} \text{[PDF] fb}
\end{align}
We can see that the process receives large NLO QCD corrections, about $89\%$, and the theoretical error is significantly reduced going from LO to NLO. Moreover, the main contribution to the theoretical uncertainty comes from the scale dependence, which is obtained using the standard $7$-point scale variation. The dynamical scale yields similar results
\begin{align}
\sigma^{\text{LO}} & = 6.813^{+4.338(64\%)}_{-2.481(36\%)}\text{[scales] fb} \\
\sigma^{\text{NLO}} & = 13.22^{+2.66(20\%)}_{-2.95(22\%)}\text{[scales]}^{+0.19(1\%)}_{-0.19(1\%)} \text{[PDF] fb}
\end{align}
In this case, the NLO QCD corrections are about $94\%$. We noticed that, imposing a jet veto on the extra radiation, these corrections are dramatically reduced. Indeed, imposing a jet veto of $p_T^{\text{veto}}(j) = 50$ GeV, we have only $11\%$ NLO QCD corrections for the fixed scale choice, $23\%$ for the dynamical one. This suggests that the hard extra radiation carries the NLO QCD contribution.

To reproduce the results available in literature \cite{Denner:2020orv}, we changed our scale prescription to
\begin{equation}
\mu_0 = \mu_{\text{DLP}} = \frac{1}{2}\biggl[\biggl(p_T^{miss} + \sum_{i = e^+,\mu^-,b_1,b_2,b_3,b_4,j} E_T(i)\biggr) + 2m_t \biggr]^{1/2} \biggl( \sum_{i = b_1,b_2,b_3,b_4,j} E_T(i)\biggr)^{1/2}
\end{equation}
We found very good agreement both at the integrated
\begin{align}
\sigma^{\text{LO}}_{\text{HELAC-NLO}} = 5.201(2)^{+60\%}_{-35\%} \text{ fb} & \spazio \sigma^{\text{LO}}_{\text{DLP}} = 5.198(4)^{+60\%}_{-35\%} \text{ fb} \\
\sigma^{\text{NLO}}_{\text{HELAC-NLO}} = 10.28(1)^{+18\%}_{-21\%} \text{ fb} & \spazio \sigma^{\text{NLO}}_{\text{DLP}} = 10.28(8)^{+18\%}_{-21\%} \text{ fb}
\end{align}
and at the differential level. These results confirm the validity of both calculations.

Finally, we compared our predictions to the experimental results obtained by ATLAS \cite{ATLAS:2018fwl}. We adapted to their cuts
\begin{align}
p_T(\ell) > 25~\text{GeV}, \hspazio |y(\ell)|  < 2.5, & \hspazio p_T(b) > 25~\text{GeV}, \hspazio |y(b)| < 2.5, \nonumber \\
\Delta R(bb) > 0.4, & \hspazio \Delta R(\ell b) > 0.4.
\end{align}
and we used our default dynamical scale. The final state they are considering requires exactly one electron and one muon (with opposite charges). We will label this final state $e\mu + 4b$. Therefore, we need to multiply by a factor of $2$ our prediction. Moreover, they include also leptonic $\tau$ decays. Assuming the following branching ratios $\mathcal{BR}(\tau^- \rightarrow \mu^- \bar{\nu}_\mu \nu_\tau ) =17.39\%$ and $\mathcal{BR}(\tau^- \rightarrow e^- \bar{\nu}_e \nu_\tau ) =17.82\%$ \cite{br}, we estimate a contribution to our theoretical prediction of $0.6$ fb. Thus, our final theoretical prediction is given by
\begin{equation}
\sigma_{e\mu+4b}^{\text{HELAC-NLO}} = (20.0 \pm 4.3) \text{ fb},
\end{equation}
which is in very good agreement with the experimental result
\begin{equation}
\sigma_{e\mu+4b}^{\text{ATLAS}} = (25 \pm 6.5) \text{ fb}.
\end{equation}
In the following table you can see a comparison of various theoretical predictions, taken from \cite{ATLAS:2018fwl} and generated with several Monte Carlo frameworks \cite{Cascioli:2013era, Alwall:2014hca, Kardos:2013vxa, Garzelli:2014aba, Bevilacqua:2017cru}, to our prediction and to the experimental measurement:
\begin{center}
\begin{tabular}{lc}
\toprule
Theoretical predictions & $\sigma_{e\mu+4b}$ [fb] \\
\midrule
\textsc{Sherpa+OpenLoops} (4FS) & $17.2 \pm 4.2$ \\
\textsc{Powheg-Box+Pythia 8} (4FS) & $16.5$ \\
\textsc{PowHel+Pythia 8} (5FS) & $18.7$ \\
\textsc{PowHel+Pythia 8} (4FS) & $18.2$ \\
\midrule
\textsc{Helac-Nlo} (5FS) & $20.0 \pm 4.3$ \\
\midrule
Experimental result (\textsc{Atlas}) & $25 \pm 6.5$ \\
\bottomrule
\end{tabular}
\end{center}
This table shows that our prediction is the closest to the experimental result. The other predictions have been obtained matching fixed order calculations to parton shower algorithms. A comparison between the full off-shell calculation and those matched to the parton shower should be performed to understand the source of this discrepancy.

  \section{Differential distributions}
  \label{sec:diff}
\begin{figure}
\centering
{\includegraphics[width=0.55\textwidth]{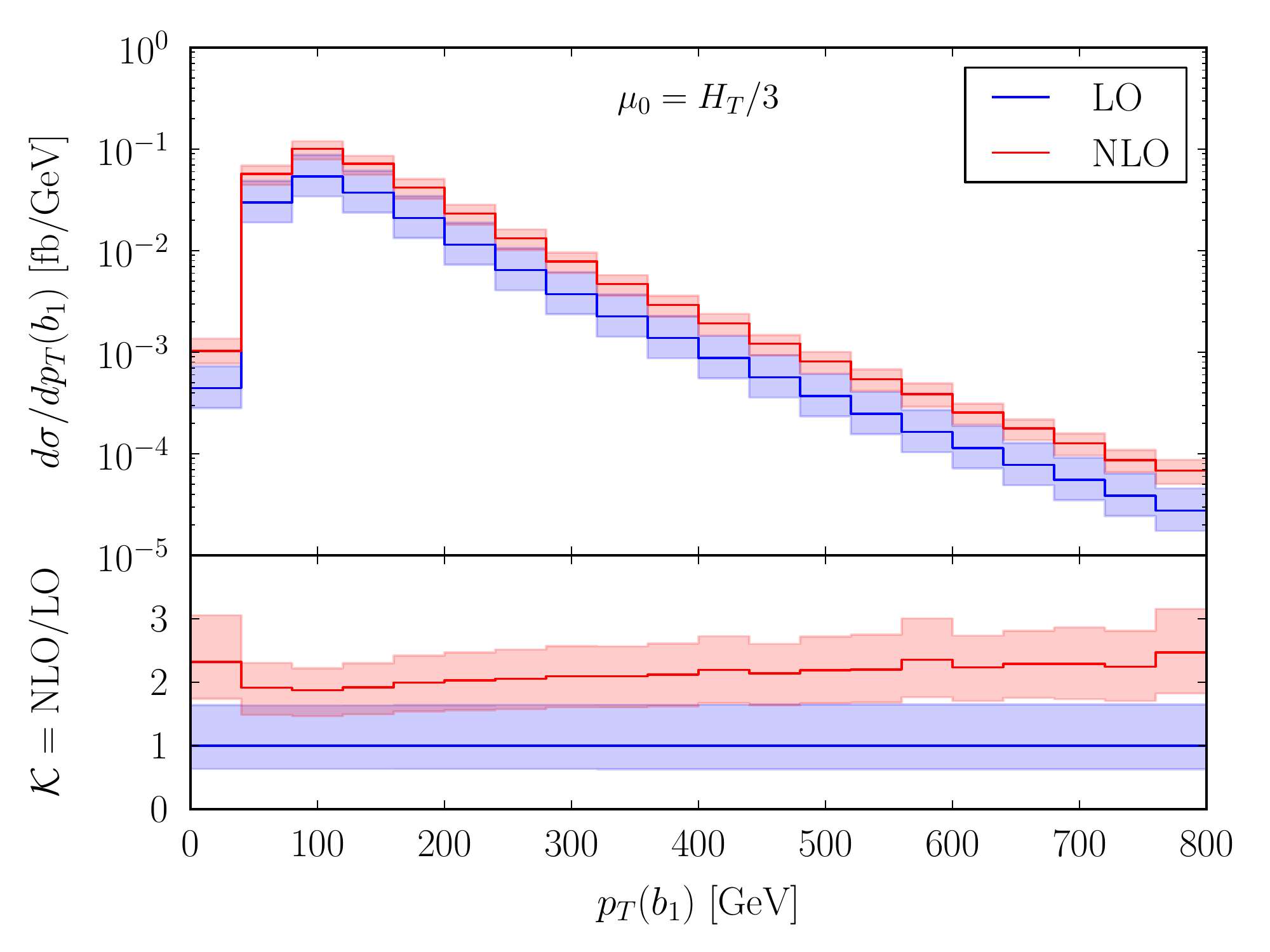}}
{\includegraphics[width=0.35\textwidth]{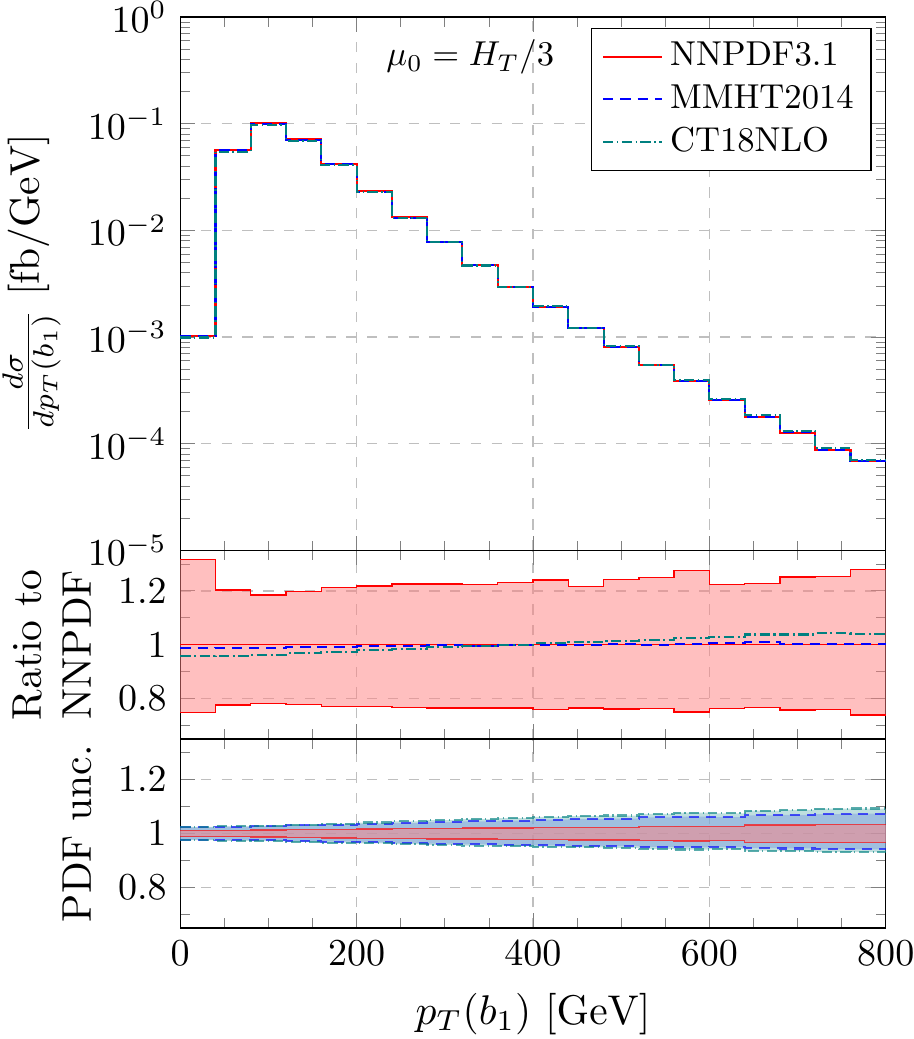}}
\caption{Differential cross section distribution as a function of the transverse momentum of the hardest $b$-jet. The left plot displays the LO and NLO distributions with the respective scale dependence. The right plot shows the uncertainties coming from the various PDF set choices.}
\label{fig:ptb1}
\end{figure}

  The higher order corrections can also affect kinematic distributions. To asses the size of these effects, we investigated differential distribution for several observables. Here we present the differential distribution of the hardest $b$-jet (Fig.~\ref{fig:ptb1}).
We found significant shape changes going from LO to NLO, with corrections from $90\%$ to $135\%$, yielding a shape distortion of $45\%$. Similarly to the integrated result, the main source of theoretical error comes from the scale dependence, which at NLO is reduced to $20-30\%$. To obtain the results for the differential cross sections we used the dynamical scale $H_T/3$. Indeed, the fixed scale works poorly in the high energy kinematic region, where the perturbative stability is spoiled. This also yields huge shape distortions that can reach $150\%$. To summarise, these results confirm that NLO QCD effects are very important to this process.
  
  \section{Contribution of initial state $b$-quarks}
  \label{sec:inib}
  Despite using the $5$ flavour scheme, until this point we ignored the initial states involving $b$-quarks, because we expect them to be negligible. In the following, we are going to show that this is actually the case. To perform this study, we first introduce two $b$-jet tagging schemes: the \textit{charge aware} and the \textit{charge blind} tagging schemes. We employ the anti-$k_t$ algorithm with $R=0.4$ to cluster the partons into jets.

The \textit{charge aware tagging scheme} is sensitive to the flavour and the charge of the $b$-jet. The recombination rules are
\begin{equation}
b\bar{b} \rightarrow g,  \hspazio bb \rightarrow b, \hspazio \bar{b}\bar{b} \rightarrow \bar{b}, \hspazio bg \rightarrow b, \hspazio \hspazio \bar{b}g \rightarrow \bar{b}.
\end{equation}
We need now to take into account the following initial states: $b\bar{b}$, $bg$ and $\bar{b}g$. In the bottom-gluon channel we can now have up to $5$ $b$-jets. Therefore, we require our final state to have at least $2$ $b$-jets and $2$ $\bar{b}$-jets. The advantage of this jet algorithm is of course that we can distinguish between $b$- and $\bar{b}$-jets. The drawback is that, from an experimental point of view, this might reduce the $b$-jet tagging efficiency, leading to smaller event statistics.

The \textit{charge blind tagging scheme} is sensitive to the absolute flavour and does not attempt to tag the charge of the $b$-jet. The recombination rules are
\begin{equation}
b\bar{b} \rightarrow g,  \hspazio bb \rightarrow g, \hspazio \bar{b}\bar{b} \rightarrow g, \hspazio bg \rightarrow b, \hspazio \hspazio \bar{b}g \rightarrow \bar{b}.
\end{equation}
Therefore, we require our final state to have at least $4$ $b$-jets, independently of their charge. Because we do not keep track of the charge of the $b$-jets, we need to consider in addition the following initial states: $bb$ and $\bar{b}\bar{b}$. In contrast to the charge aware tagging scheme, the $b$-jet tagging efficiency will be better, leading to larger event statistics, but the price to pay is the loss of the information on the charge.

\begin{figure}
\centering
{\includegraphics[width=0.48\textwidth]{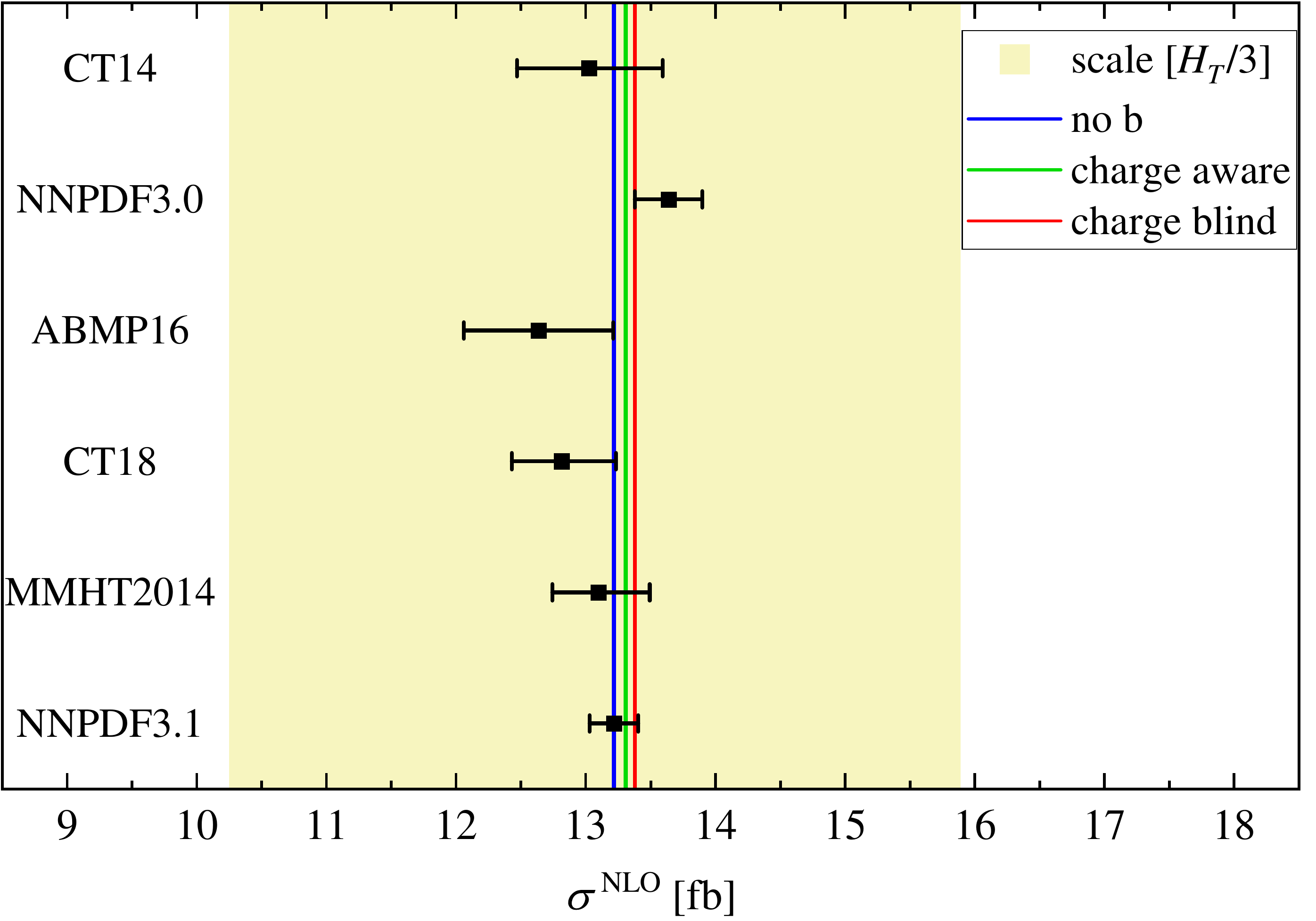}}
{\includegraphics[width=0.42\textwidth]{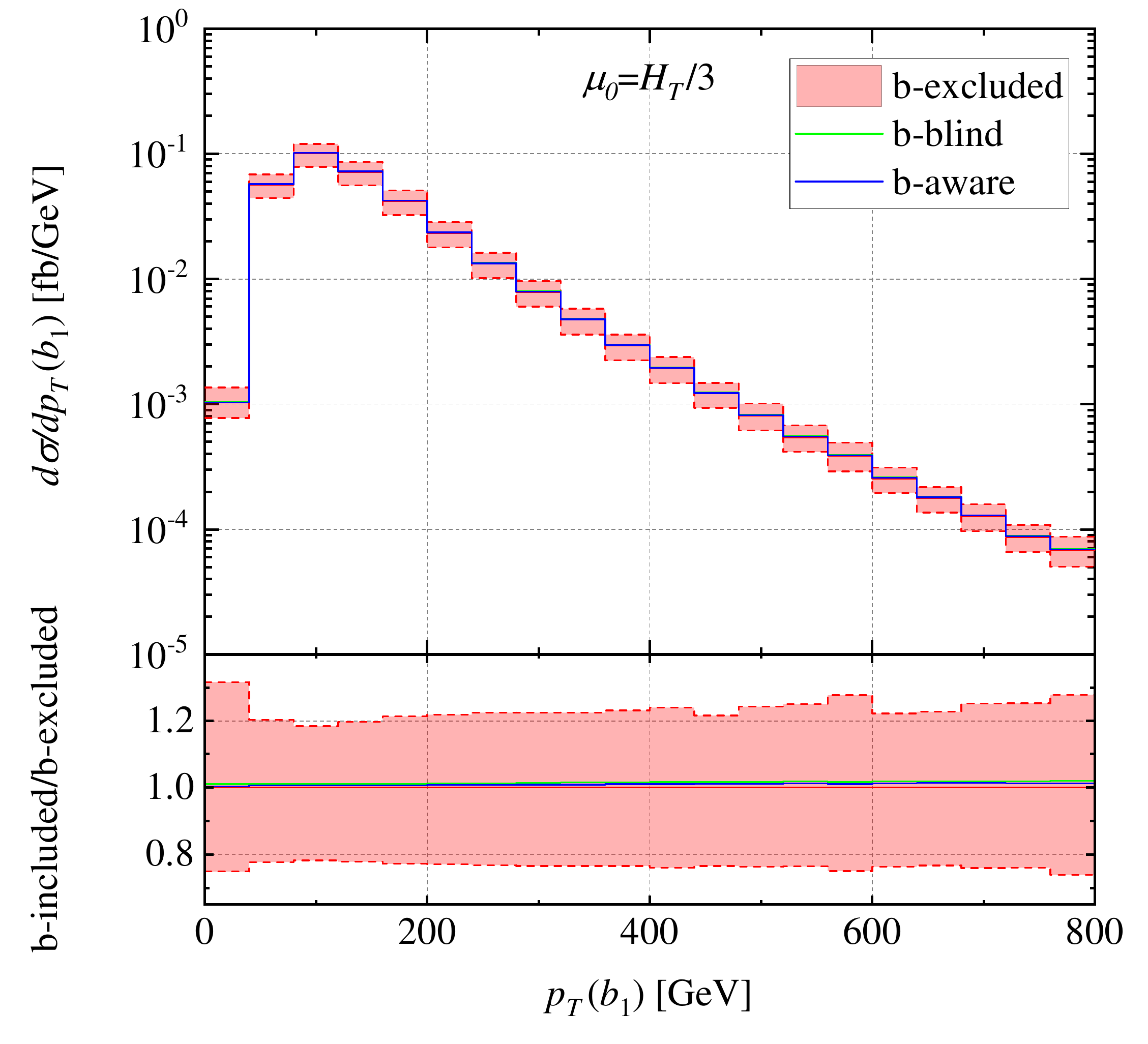}}
\caption{Comparison of the predictions including and excluding $b$-quarks from the initial state. The left plot reports the results for the integrated fiducial cross section, where also scale and pdf uncertainties are displayed. The right plot shows the comparison at the differential level for the transverse momentum distribution of the hardest $b$-jet.}
\label{fig:inib}
\end{figure}

In the following we report the results obtained using the two tagging schemes and we compare them to the results where the initial states involving $b$-quarks are neglected. We used our standard setup and the dynamical scale choice $H_T/3$. The LO integrated fiducial cross sections are
\begin{equation}
\sigma_{\text{no b}}^{\text{LO}} = 6.813(3) \text{fb}, \hspazio \sigma_{\text{aware}}^{\text{LO}} = 6.822(3) \text{fb}, \hspazio \sigma_{\text{blind}}^{\text{LO}} = 6.828(3) \text{fb}.
\end{equation}
We can see that the effects of these initial state contributions are up to $0.2\%$. The situation at NLO is not much different
\begin{equation}
\sigma_{\text{no b}}^{\text{NLO}} = 13.22(3) \text{fb}, \hspazio \sigma_{\text{aware}}^{\text{NLO}} = 13.31(3) \text{fb}, \hspazio \sigma_{\text{blind}}^{\text{NLO}} = 13.38(3) \text{fb}.
\end{equation}
The effects are now up to $1\%$, well within the theoretical uncertainty, as we can see from Fig.~\ref{fig:inib}. We could not see relevant effect neither at the differential level. We can conclude that we can safely neglect $b$-quark initial states in our process.

  \section{Summary and Outlook}
  \label{sec:summ}
  We presented LO and NLO full off-shell predictions for $pp \rightarrow e^+ \nu_e b \mu^- \bar{\nu}_\mu \bar{b} b \bar{b} + X$. We observed huge NLO QCD corrections both at the integrated and differential level. These can be reduced applying a jet veto to the extra radiation. The theoretical uncertainties drop going from LO to NLO, where they are about $20\%$ and mainly come from the scale dependence. Our predictions are in very good agreement both with previous results \cite{Denner:2020orv} and with experimental measurements \cite{ATLAS:2018fwl}. We finally investigated the contribution of the initial states involving $b$-quarks, which turns out to be negligible. To do so, we introduced two $b$-jet tagging schemes that we called \textit{charge aware} and \textit{charge blind} tagging schemes.

In the future we want to asses the size of the off-shell effects by studying the very same process using the so-called Narrow Width Approximation, and comparing it to the full off-shell calculation. Moreover, we aim to find a prescription to label the $b$-jets produced in the process, to distinguish between those coming from the decay of the top-quarks and the remaining ones. This would help to distinguish the background from the actual $t\bar{t}H$ signal in the $M(bb)$ distribution of the Higgs boson.

\section*{Acknowledgements}
The research of M.L. was supported by the DFG under grant 400140256 - GRK 2497: \textit{The physics
of the heaviest particles at the Large Hardon Collider}.

\bibliography{mybib}
\bibliographystyle{SciPost_bibstyle}




\nolinenumbers

\end{document}